\documentclass[12pt]{article}
\usepackage{epsfig}

\def\beq{\begin{equation}}
\def\eeq#1{\label{#1}\end{equation}}
\def\eeqn{\end{equation}}

\def\beqa{\begin{eqnarray}}
\def\eeqa#1{\label{#1}\end{eqnarray}}
\def\eeqan{\end{eqnarray}}

\let\bar=\overbar

\def\Dslash{\not{\hbox{\kern-4pt $D$}}}
\def\dslash{\not{\hbox{\kern-2pt $\del$}}}

\def\msb{{\bar{\ssstyle M \kern -1pt S}}}

\def\Title#1{\begin{center} {\Large {\bf #1} } \end{center}}

\begin{document}

\Title{
Two Top Utilities of Two Higgs Doublets:
 Electroweak Baryogenesis \& Alignment\footnote{
Talk presented at the APS Division of Particles and Fields Meeting
(DPF 2017), July 31-August 4, 2017, Fermilab. C170731}
}

\bigskip\bigskip


\begin{raggedright}

{\it George W.-S. Hou\footnote{
Home institute: Department of Physics, National Taiwan University,
Taipei 10617, Taiwan.} \\
ARC CoEPP at the Terascale \\
School of Physics, University of Melbourne\\
Melbourne, Vic 3010, AUSTRALIA
}
\bigskip\bigskip
\end{raggedright}

\section{Introduction 
}

The Standard Model (SM) 
falls short of electroweak baryogenesis (EWBG) 
on order of electroweak phase transition (EWPT)
and strength of $CP$ violation (CPV): 
weak interaction is too weak, while the 
Jarlskog invariant is overly suppressed by quark masses and mixings.
Adding a second Higgs doublet (2HDM), 
one could have 1st order EWPT if Higgs quartic couplings are ${\cal O}(1)$.
%
Since known CPV come from the CKM matrix, i.e. Yukawa couplings,
can there be extra Yukawa couplings in 2HDMs?
Such couplings were killed by the Natural Flavor Conservation
(NFC) condition of Glashow and Weinberg~\cite{Glashow:1976nt},
usually by imposing $Z_2$ symmetry on the Higgs fields
to forbid flavor-changing neutral Higgs (FCNH) couplings.
But such discrete symmetries are \emph{ad hoc}, and given 
the observed trickle-down pattern or mass suppression of off-diagonal quark mixings, 
it was deemed~\cite{Cheng:1987rs} perhaps an overkill.

We drop $Z_2$ symmetry (or NFC) and utilize extra Yukawa couplings
$\rho_{tt}$ and $\rho_{tc}$, naturally ${\cal O}(1)$ and complex,
to drive~\cite{Fuyuto:2017ewj} EWBG.
Note that 
many authors have recently taken a data-driven approach to FCNH couplings,
not just for $t\to ch$ decay~\cite{Hou:1991un}, 
but applying also to the $B \to D^{(*)}\tau\nu$ anomaly 
and $h \to \tau\mu$ decay~\cite{Chang:2017wpl}.

As ${\cal O}(1)$ Higgs sector couplings is a prerequisite for EWBG,
we find an interesting second utility~\cite{Hou:2017hiw}:
 \emph{bringing about alignment naturally.}
The fact that the observed 125 GeV $h^0$ boson 
resembles closely~\cite{Khachatryan:2016vau} the SM-Higgs, i.e. alignment,
poses an embarrassment for supersymmetry, as well as the
associated 2HDM II that require $u$- and $d$-type quarks 
to receive mass from \emph{separate} scalar doublets ($Z_2$ symmetry).
One needs to invoke decoupling~\cite{Gunion:2002zf} and send the exotic Higgs 
$\Phi'$ to multi-TeV, which seems rather high compared to 125 GeV,
and out of reach at the LHC.
In the second half of this talk,
we show that alignment, or small mixing between the two $CP$-even scalars,
emerges naturally in 2HDM without $Z_2$
with ${\cal O}(1)$ Higgs quartic couplings.

The sub-TeV exotic Higgs bosons should be 
a boon to LHC search.

\section{\boldmath 2HDM without $Z_2$:  FCNH $\rho_{ij}$ Couplings}

The Yukawa interaction for up-type quarks
in 2HDM without $Z_2$  is
\begin{eqnarray}
-\mathcal{L}_Y
 = \bar{q}_{iL}\left(Y^u_{1ij}\tilde{\Phi}_1+Y^u_{2ij}\tilde{\Phi}_2 \right) u_{jR}+{\rm h.c.},
\end{eqnarray}
where $i, j$ are flavor indices and $\tilde \Phi_b = i\tau_2\Phi^*_b$ ($b=1, 2$).
With $\Phi_{1,2}$ each acquiring a vacuum expectation value (VEV) $\upsilon_{1,2}$,
and defining as usual $\upsilon_1 = \upsilon c_\beta$, $\upsilon_2 =\upsilon s_\beta$
($\upsilon^2 = \upsilon_1^2 + \upsilon_2^2$),
the matrix $Y^{\rm SM}= Y_1\, c_{\beta}+Y_2\, s_{\beta}$
is diagonalized by $V_L^{u\dagger}Y^{\rm SM}V^u_R$ to $Y_D$,
with diagonal elements $y_i \equiv \sqrt2 m_i/\upsilon$.
However, the orthogonal combination
\begin{eqnarray}
\rho = V_L^{u\dagger}\left(-Y_1\, s_{\beta}+Y_2\, c_{\beta} \right)V^u_R,
\end{eqnarray}
cannot be simultaneously diagonalized,
and the exotic neutral Higgs bosons $H^0$ and $A^0$ possess
FCNH couplings in general, including extra diagonal couplings $\rho_{ii}$,
\begin{equation}
-\sqrt{2}\mathcal{L}_Y
 = \bar{u}_{iL}\left[
   ({y_i\delta_{ij}}\, s_{\beta-\alpha} + {\rho_{ij}}\, c_{\beta-\alpha})\, h 
 + ({y_i\delta_{ij}}\, c_{\beta-\alpha} - {\rho_{ij}}\, s_{\beta-\alpha})\, H 
 -{i}\, \rho_{ij} \,\gamma_5\, A\right] u_{jR} + {\rm h.c.}
\end{equation}
The new Yukawa couplings $\rho_{ij} = |\rho_{ij}| e^{\phi_{ij}}$ are complex,
and $c_{\beta-\alpha}$ is the mixing angle between $h^0$ and $H^0$.
The discovered $h^0$ is rather close to
SM-Higgs, i.e. we are close to the \emph{alignment} limit~\cite{Gunion:2002zf}
of $c_{\beta-\alpha} \to 0$ (hence $|s_{\beta-\alpha}| \to 1$).
In this limit, the Yukawa couplings of $h^0$ are diagonal,
while $H^0$ and $A^0$ have FCNH couplings $\rho_{ij}$.
%

We note that $\tan\beta = v_2/v_1$ is unphysical at zero temperature.

\section{\boldmath EWBG driven by $\rho_{tt}$}

Let us first give a heuristic account of EWBG.

Baryon number violation is facilitated by sphaleron processes in the symmetric phase.
As $T$ drops, one has an expanding bubble of the broken phase.
But to avoid washout of the baryon number $n_B$ across the bubble wall, 
one needs $\Gamma_B^{\rm{(br)}}(T_C)<H(T_C)$,
i.e. the $n_B$ changing rate $\Gamma_B^{(\rm{br})}(T_C)$ is less than
the Hubble parameter $H(T_C)$ at critical temperature $T_C$.
This can be satisfied if EWPT is first order such that
$\upsilon_C/T_C > 1$, where $\upsilon_C^2 = \upsilon_1^2(T_C) + \upsilon_2^2(T_C)$.
A strongly 1st order EWPT can be achieved in 2HDM through thermal loops 
involving extra Higgs bosons with ${\cal O}(1)$ quartic couplings, 
in contrast to the rather weak Higgs self-coupling in SM.

The baryon asymmetry of the Universe (BAU), or $n_B/s \equiv Y_B \neq 0$, arises via
\begin{equation}
Y_B \equiv \frac{n_B}{s} =
 \frac{-3\Gamma_B^{(\rm{sym})}}{2D_q \lambda_+ s}
 \int_{-\infty}^0 dz' n_L(z') e^{-\lambda_-z'},
\end{equation}
where
 $\Gamma_B^{(\rm{sym})} = 120\alpha_W^5T$ is the $n_B$-changing rate in symmetric phase,
 $D_q \simeq 8.9/T$ is the quark diffusion constant,
 $s$ is the entropy density,
 $\lambda_\pm \sim \upsilon_w$ is the bubble wall velocity, and
 $n_L$ is the total left-handed fermion number density.
The integration is over $z'$, the coordinate opposite bubble expansion direction.
We use the Planck value $Y_B^{\rm obs}= 8.59\times 10^{-11}$~\cite{Ade:2013zuv}
in our numerical analysis.

\begin{figure}[t]
\center
\includegraphics[width=3.2cm]{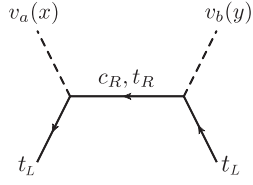}
\caption{
Leading CPV process for BAU, where $\upsilon_{a}(x)$ and $\upsilon_b(y)$ 
denote bubble wall.
}
\label{fig:bubble_int}
\end{figure}

\vskip0.2cm
\noindent\underline{CPV Top Interactions}
\vskip0.1cm

Nonvanishing $n_L$ is needed for $Y_B$, which in our case is 
from the l.h. top density.
CPV interactions of (anti)top with the bubble wall
is illustrated symbolically in Fig.~1,
where vertices can be read off from Eq.~(1).
The detailed ``transport'' equations are rather elaborate,
which we do not go into detail here.
Suffice it to say that,
with the closed time path formalism in the VEV insertion approximation,
the CPV source term $S_{ij}$ for l.h. fermion $f_{iL}$
induced by r.h. fermion $f_{jR}$ takes the form
\begin{equation}
{S_{i_L j_R}(Z)}= N_C F\,
    \rm{Im}\big[(Y_1)_{ij}(Y_2)_{ij}^*\big]\, v^2(Z)\, \partial_{t_Z}\beta(Z),
\end{equation}
where
 $Z = (t_Z,0,0,z)$ is position in heat bath (very early Universe),
 $N_C = 3$ is number of color,
 $F$ is a function\footnote{
    See Ref.~\cite{Chiang:2016vgf} for explicit form,
    as well as more details on the transport equations.}
  of complex and $T$-dep. energies of $f_{iL}$ and $f_{jR}$,
 and $\partial_{t_Z}\beta(Z)$ is the variation in $\beta(Z)$.
Note that, even though $\beta$ is basis-dependent in 2HDM without $Z_2$,
its variation is physical and
plays an essential role in generating the CPV source term.
In our numerics, we take $\Delta\beta = 0.015$.

If bubble wall expansion and $\partial_{t_Z}\beta(Z)$
reflect departure from equilibrium, CPV for BAU is in the
$\rm{Im}\big[(Y_1)_{ij}(Y_2)_{ij}^*\big]$ factor in Eq.~(5).
Let us see how it depends on the $\rho_{ij}$ couplings.
From Eq. (2) and the relation between $Y^{\rm SM}$ and $Y_D$,
one has
\begin{equation}
\rm{Im}\big[(Y_1)_{ij}(Y_2)_{ij}^*\big]
= \rm{Im}\big[(V_L^uY_{\rm diag} V_R^{u\dagger})_{ij}(V_L^u\rho V_R^{u\dagger})_{ij}^* \big].
\end{equation}
To appreciate the results presented below,
suppose~\cite{Guo:2016ixx} $(Y_{1})_{tc} \neq 0$, $(Y_{2})_{tc} \neq 0$,
and $(Y_1)_{tt}=(Y_2)_{tt} \neq 0$, and all else vanish
(taking $\tan\beta =1$ throughout for convenience).
Then $\sqrt{2}Y^{\rm SM} = Y_1 + Y_2$ can be
diagonalized by just $V_R^u$ to a single nonvanishing
$33$ element $y_t$, the SM Yukawa coupling,
while $-Y_1 + Y_2$ is not diagonalized.
Solving for $V_R^u$ in terms of nonvanishing elements
in $Y_1$ and $Y_2$, one finds
\begin{equation}
  \rm{Im}\big[ (Y_1)_{tc}(Y_2)_{tc}^* \big]
 = -y_t\, \rm{Im}(\rho_{tt}), \quad \rho_{ct}=0,
\end{equation}
with $\rho_{tc}$ basically a free parameter.
Note that both doublets are involved for EWBG.

To calculate $n_L$ in Eq.~(4), one has 
a set of diffusion equations fed by various density combinations 
weighted by mass-dep. statistical factors, 
as well as CPV source terms such as Eq.~(5).
By a relatively standard treatment, the set of coupled equations 
is reduced to a single one for $n_H$ controlled by diffusion time 
$D_H \simeq 101.9/T$ modulated by $1/\upsilon_w^2$
(see Ref.~\cite{Chiang:2016vgf} for discussion and references).
%
Important parameters are given in Ref.~\cite{Fuyuto:2017ewj},
and we note that $\upsilon_C/T_C > 1$ is satisfied.

\begin{figure}[t]
\center 
\includegraphics[width=6.7cm]{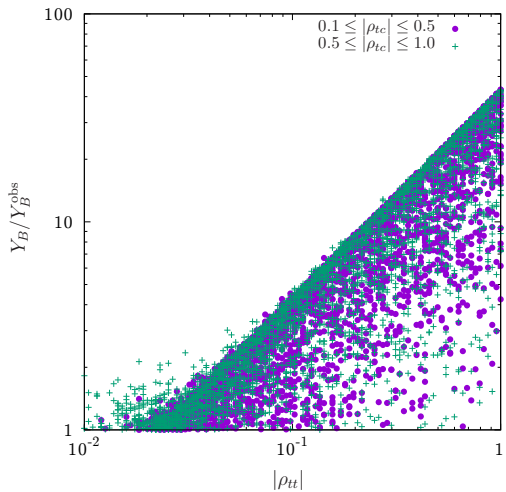} 
\caption{
$Y_B/Y_B^{\rm obs}$ vs $|\rho_{tt}|$ with purple dot (green cross)
 for $|\rho_{tc}| \in [0.1,\, 0.5]$ ($[0.5,\, 1]$).
}
\end{figure}

Scanning over $|\rho_{tc}|$, $\phi_{tt}$ and $\phi_{tc}$,
we plot $Y_B/Y_B^{\rm obs}$ vs $|\rho_{tt}| \in (0.01,\, 1)$ in Fig.~2,
with $\rho_{tt}$ and $\rho_{tc}$ satisfying~\cite{Altunkaynak:2015twa}
$B_d$, $B_s$ mixing as well as $b\to s\gamma$ constraints.
We have taken $m_H = m_A = m_{H^\pm} = 500$ GeV to simplify.
Though perhaps too restrictive, it illustrates the charm of EWBG:
the exotic Higgs masses are sub-TeV.
We separate $|\rho_{tc}| \in (0.1,\, 0.5)$ and $(0.5,\, 1)$,
plotted as purple dots and green crosses, respectively.
Sufficient $Y_B$ can be generated handsomely,
even for $|\rho_{tt}|$ below 0.1.
Since no obvious difference is seen for lower vs higher $|\rho_{tc}|$
for the bulk of the plot, we infer that $Y_B$ is driven by $\rho_{tt}$.
However, note that for small $|\rho_{tt}|$,
the green crosses populate $Y_B/Y_B^{\rm obs} > 1$
much more than the purple dots, which means that $|\rho_{tc}| \sim {\cal O}(1)$
could take over EWBG for low $\rho_{tt}$,
but near maximal $\phi_{tc}$ would be needed.

Thus, we have $\rho_{tt}$ as main driver for BAU, 
with $\rho_{tc}$ at ${\cal O}(1)$ as backup.

\section{Interlude: the Alignment Enigma}

Let's get back on Earth.

Nathaniel Craig gave the SUSY and BSM plenary talk at 
EPS-HEP 2017 meeting in Venice, where he placed SUSY scale at 5 TeV.
Though he talked about the scale of various SUSY particles,
he did not mention heavy Higgs.
So, at the end I asked: 
``In SUSY there is a second Higgs doublet,
where do you place them?'', to which he replied ``3 TeV'', and 
I expressed understanding that he took the decoupling limit~\cite{Gunion:2002zf}.
However,
the squared mass ratio with 125 GeV boson would be about a factor of 600,
which is rather fine a tuning.
To further illustrate fine-tuning, 
we mention a new solution found in Ref.~\cite{Carena:2013ooa} 
for alignment without decoupling in 2HDM II,
\begin{equation}
\tan\beta = \frac{\lambda_{\rm SM} - \tilde \lambda}{\lambda_7}
  + \; {\rm small\ terms}.
\end{equation}
The r.h.s. can be large because the numerator is ${\cal O}(1)$,
since $\lambda_{\rm SM} \simeq 0.26$ and $\tilde \lambda$
is a combination of Higgs quartics, 
while denominator is generated by soft breaking hence small.
The solution for large $\tan\beta$ does exist,
but the two sides are unrelated, hence it amounts to
an \emph{accidental cancellation} according to Howard Haber.

Haber elucidated this alignment issue at his Toyama talk in early 2017.
Going to Higgs basis, so only one Higgs doublet has VEV,
he gives~\cite{Bechtle:2016kui}
\begin{equation}
  \cos(\beta - \alpha) \simeq \frac{Z_6 v^2}{m_H^2 - m_h^2} \ll 1,
\end{equation}
which can be realized, he said, either by large denominator, or small $Z_6$.
With SUSY mindset, the latter is typically~\cite{Bechtle:2016kui} 0.05 or smaller.

We find one can shake off this MSSM mindset,
that alignment naturally emerges in 2HDM without $Z_2$,
i.e. if one throws away the yoke of $\tan\beta$.

\section{
\boldmath Bonus of 2HDM without $Z_2$: Alignment}

The Higgs potential of 2HDM without $Z_2$ symmetry, 
assumed to be $CP$ invariant, is
\begin{eqnarray}
 &V(\Phi,\, \Phi')
= \mu_{11}^2 |\Phi|^2 +\mu_{22}^2 |\Phi'|^2
        - \left(\mu_{12}^2 \Phi^\dagger \Phi' + {\rm h.c.}\right)
   +\frac{\eta_1}{2}|\Phi|^4 + \frac{\eta_2}{2}|\Phi'|^4 + \eta_3|\Phi|^2|\Phi'|^2 \nonumber\\
   &\ + \eta_4|\Phi^\dagger \Phi'|^2 
   + \left\{\frac{\eta_5^{}}{2}(\Phi^\dagger \Phi')^2
   + \left[\eta_6^{}|\Phi|^2 + \eta_7^{}
      |\Phi'|^2\right]\Phi^\dagger \Phi' + {\rm h.c.}\right\},
\end{eqnarray}
where $\eta_6$, $\eta_7 \neq 0$ are now allowed. 
Choosing $\Phi$ to generate $v$, i.e. in Higgs basis,
one gets besides $\mu_{11}^2 = -\frac{1}{2}\eta_1^{}v^2$
a second relation $\mu_{12}^2 =  \frac{1}{2}\eta_6^{}v^2$
that absorbs the ``soft-breaking'' $\mu_{12}^2$ parameter,
with $\mu_{22}^2$ now positive definite.

\begin{figure}[t]
\center
\includegraphics[width=7.5cm]{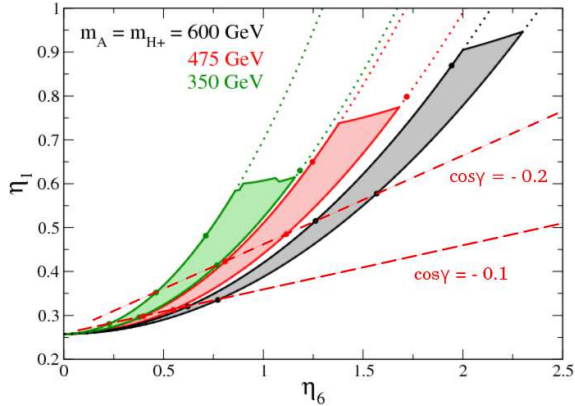}
\caption{
Range for $\eta_1$ vs $\eta_6$ for $m_A = m_{H^+} = 350$, 475, 600 GeV
 and $\eta_4 = \eta_5 \in (0.5,\; 2)$, cut off by $\Delta T$ (dotted),
where filled circles are for $-\cos\gamma = 0.1,\ 0.2,\ 0.3$.
}
\end{figure}

The $CP$-even Higgs mass matrix is
\begin{eqnarray}
  M_\textrm{even}^2 =
  \left[\begin{array}{cc}
    \eta_1^{}v^2 & \eta_6^{} v^2 \\
    \eta_6^{} v^2 & \mu_{22}^2 + \frac{1}{2} (\eta_3^{} + \eta_4^{} + \eta_5)v^2\\
    \end{array}\right],
\end{eqnarray}
which is diagonalized by
\begin{eqnarray}
  R^T_\gamma M_\textrm{even}^2 R_\gamma  =
    \left[\begin{array}{cc}
    m_H^2 & 0 \\
    0 & m_h^2 \\
  \end{array}\right], \quad
  R_\gamma  =  \left[\begin{array}{cc}
    c_\gamma & - s_\gamma \\
    s_\gamma & c_\gamma \\
  \end{array}\right],
\end{eqnarray}
where the convention is close to 2HDM II,
i.e. $c_\gamma \equiv \cos\gamma$ replaces $\cos(\beta -\alpha)$,
and $s_\gamma \equiv \sin\gamma$.
Close to alignment limit, $c_\gamma \to 0$, $s_\gamma \to -1$,
one finds, similar to Eq.~(9)
\begin{eqnarray}
  c_\gamma^{} & \cong \frac{-\eta_6^{}v^2}{m_H^2 - m_h^2}.
\end{eqnarray}
Note that $\eta_6$ controls the mixing, 
while $\eta_1$ is less important and $\eta_7$ does not enter.

Unlike Ref.~\cite{Bechtle:2016kui}, we find in general
\emph{there is no need for small $\eta_6$!} 
For $c_\gamma$ to be small, we find the rough condition
\begin{eqnarray}
  |\eta_6| < \eta_{3,\,4,\,5}
   \sim {\cal O}(1) < \mu_{22}^2/v^2, 
\end{eqnarray}
and $\eta_{3,\,4,\,5} > 0$, i.e. positivity, would help.
The value of $m_h$ need not be as small as 125 GeV.
However, given that $m_h^2/v^2 \simeq 0.26$,
taking this value for $\eta_6$ would give rather small $c_\gamma$. 
The gist of it is that, both 11 and 12 (21) entries of Eq.~(11)
have just one term, while the 22 entry has 4, due to the enlarged Higgs sector.
Thus, 1/4, close to the Cabibbo angle, is the
natural starting point for \emph{emergent} alignment.

For numerical illustration,
we need to take into account electroweak $\Delta T$ constraint.
To this end, we take $\eta_4 = \eta_5$ hence $m_A = m_{H^+}$, 
which is the custodial SU(2) that eliminates the scalar-scalar self-energy,
hence $\Delta T$ arises from a scalar-vector term.
Though suppressed by $M_Z^2 - M_W^2$,
we will see that $\Delta T$ still puts a constraint.
We plot the range of $\eta_1$ vs $\eta_6$ in Fig.~3, where
$-c_\gamma^{} = 0.1,\ 0.2,\ 0.3$ ($-\sin\gamma \cong 0.995,\ 0.980,\ 0.954$)
are marked by filled circles.
We see that, for $m_H =350$, 475, 600 GeV,
one has $-c_\gamma < 0.2$ for $\eta_6 < 0.5$, 1, 1.5,
respectively, i.e. close to alignment,
while for $\eta_6 \sim m_h^2/v^2 \cong 0.26$,
one is \emph{very} close to the alignment limit.
This is in strong contrast to the tuning of 2HDM~II
reflected in Eq.~(8):
\emph{alignment is natural in 2HDM without $Z_2$ symmetry}.
We note further that, while $\eta_1 < \eta_6$ in general,
there is large parameter space where both are ${\cal O}(1)$,
but $m_h^2$ is pushed down from $\eta_1 v^2$ with help of level splitting.

We comment that the one-loop $\rho_{tt}$ effect on $h \to ZZ^*$, 
with $\rho_{tt}c_\gamma < 0$, can compensate~\cite{Hou:2017vvp} 
the general suppression from the bosonic loop,
which is sizable because of ${\cal O}(1)$ couplings.
This ``protection'', or ability for $\rho_{tt}$ 
to bring $h \to ZZ^*$ more in line with SM expectation,
could be reason behind the ``apparent'' alignment,
and was the original motivation to study alignment in 2HDM without $Z_2$.

\begin{figure}[t]
\center 
\includegraphics[width=6.7cm]{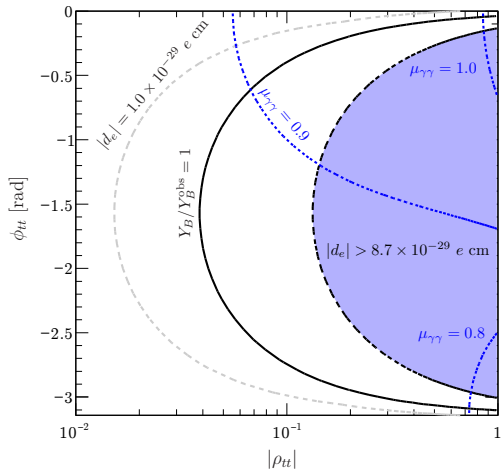}
\caption{
$Y_B/Y_B^{\rm{obs}}=1$ (solid) and $|d_e|$ (dashed)
 in the $|\rho_{tt}|$--$\phi_{tt}$ plane for $c_{\gamma} = 0.1$.
 Shaded region is excluded,
 while dotted curves are for $h \to \gamma\gamma$
 with $\mu_{\gamma\gamma}$ as marked.
}
\end{figure}

\section{Phenomenology and Discussion}

A prime FCNH effect of interest is $t\to ch$ decay,
where the latest ATLAS bound~\cite{Aaboud:2017mfd} 
using 36.1 fb$^{-1}$ data at 13 TeV gives $0.22\%$.
With $c_{\gamma}$ small but not fully known, 
$|\rho_{tc}| \sim 1$ is still allowed.
Another motivation for FCNH was a hint for $h \to \tau\mu$ 
in 8 TeV data by CMS, which has since disappeared with more data.
The new CMS bound~\cite{CMS:2017onh}, based on 35.9 fb$^{-1}$ 
at 13 TeV, gives ${\cal B}(h \to \tau\mu) < 0.25\%$.
Taking $c_{\gamma} = 0.1$, this still allows
${\cal B}(\tau \to \mu\gamma)$ up to $10^{-8}$~\cite{Fuyuto:2017ewj},
which can be probed by Belle~II.

The ACME experiment~\cite{Baron:2013eja} has made impressive 
recent progress on electron edm, $d_e$, which may correlate with EWBG.
The bound~\cite{Fuyuto:2017ewj} is shown as the dashed curve in Fig.~4, taking $c_{\gamma} = 0.1$.
Exclusion (shaded) is to its right,
with $Y_B/Y_B^{\rm{obs}}=1$ given as the solid curve.
ACME projects an improvement by factor of 9
(gray dashed curve), which would explore fully our EWBG mechanism.
However, the $d_e$ bound given in Fig.~4 is via $\rho_{tt}$ 
through the two-loop mechanism, \emph{assuming} $\rho_{ee} = 0$.
If one takes $|\rho_{ee}| \sim y_e = \sqrt2 m_e/\upsilon$ but purely imaginary,
cancellation between one- and two-loop effects could occur,
and can evade ACME. Or, ACME can probe $\rho_{ee}$!

It is important to note that the flavor or $CP$ violating effects 
mentioned so far all vanish with $c_{\gamma} \to 0$, i.e. alignment.
\emph{Alignment provides a protection that replaces NFC!}~\cite{Hou:2017hiw}
What does not vanish with $c_{\beta-\alpha}$ is EWBG itself:
Nature seems skilled at producing the Universe, but
hides the traces of flavor and CPV.

Also plotted~\cite{Fuyuto:2017ewj} in Fig.~4 are possible reductions\footnote{
Loop effect from top via $\rho_{tt}$ could compensate~\cite{Hou:2017vvp} this reduction,
which vanishes with alignment.
}
to $h \to \gamma\gamma$ width (dotted curves) due to $H^+$ effect,
which does not vanish with $c_{\gamma}$.
Another effect that survives alignment is
extra Higgs correction to $\lambda_{hhh}$, or triple-$h$ coupling,
which could receive 60\% enhancement with the
$m_H = m_A = m_{H^\pm} = 500$ GeV benchmark.

\vskip0.2cm
\noindent\underline{Niche for Extra Higgses}
\vskip0.1cm

The ``charm of EWBG", as mentioned, is of sub-TeV exotic scalars,
which can be probed directly at LHC.
This is a consequence of ${\cal O}(1)$ self-couplings in the Higgs sector.
Of course, full degeneracy is clearly too restrictive,
and the actual parameter space should be much broader,
for example, the custodial case of $m_A = m_{H^\pm}$~\cite{Hou:2017hiw}.
Knowing that $H^0$ and $A^0$ detection may be hampered 
by interference effects in $t\bar t$ decay final state,
search strategy for heavy Higgs should be readjusted.

At first issue is the bound~\cite{Misiak:2017bgg} $m_{H^+} > 570$ GeV,
based on improved Belle analysis of $B \to X_s + \gamma$.
However, unlike 2HDM II, the new bound does 
not apply to 2HDM without $Z_2$:
$b \to s\gamma$ rate now depends~\cite{Altunkaynak:2015twa} 
on $\rho_{tt}$, $\rho_{tc}$, $\rho_{cc}$ and $\rho_{bb}$, 
as well as $m_{H^+}$, and would carve out a solution space.
A targeted study is under way~\cite{Masaya}.

As mentioned, the parameter space for Higgs mass is 
much larger~\cite{Hou:2017hiw} than illustrated so far.
Positivity is not necessary.
Custodial symmetry is not necessary (it could even be twisted custodial),
and in any case, $H^0$ could be lighter than $A^0$, $H^+$.
It would be nice to link with EWBG, but it is not necessary.
As large $\mu_{22}^2$ bridges to the decoupling limit, 
however, sub-TeV spectrum is preferred.
Our mass range for illustration, $m_A = m_{H^+} = 350$, 475, 600 GeV, 
were judicially chosen.
In our custodial limit, $m_A = m_{H^+} = \mu_{22}^2 + \frac{1}{2}\eta_3v^2$.
Keeping perturbativity, say $\eta_3 < 3$, then
by 600~GeV, $\mu_{22}^2/v^2 \sim 4.5$,
and is starting to move towards decoupling,
and could e.g. damp out the first order EWPT, 
which would quench EWBG.

We see that there is a vast sub-TeV parameter space 
for ATLAS and CMS to explore at LHC.
We do not know the spectrum, which makes exploring 
$H^0$, $A^0$, $H^+$ production and decay both rich, and difficult.
There is the difficulty~\cite{Carena:2016npr} of
$gg \to H^0$, $A^0 \to t\bar t$ search, due to 
interference with $gg \to t\bar t$ background.
It should be clear that $H^+ \to t\bar b$ would be
even more hampered by ample $b$ production in association with top.
Ref.~\cite{Altunkaynak:2015twa} advocated 
$gg \to H^0$, $A^0 \to t\bar c$ search, 
which is apparently rather promising.
But the cross section is no greater than $s$-channel
single top, or $u\bar d \to t\bar b$ production,
and may get hampered by $t + j$ mass resolution.
A promising suggestion~\cite{KMH} is to capitalize on sizable $\rho_{tc}$
to search for the $cg \to tH^0$, $tA^0$ associated production,
which could feed both same sign dilepton and tri-top ($tt\bar t$) signatures.

Freeing from discrete $Z_2$ symmetry, one not only regains
the sub-TeV domain for exotic Higgs bosons from a second doublet,
the search strategy should also change.
For example, gone are the $H^+ \to \tau^+\nu$ searches,
as these are likely suppressed by $\sim y_\tau$, Yukawa of $\tau$ in SM.
But of course, $h^0 \to \tau\mu$ should continue to be pushed.

\section{\boldmath Conclusion: $H^0$, $A^0$, $H^+$ in Our Time}

In 2HDM without discrete $Z_2$ symmetry, 
we find EWBG is surprisingly efficient with the
combination of ${\cal O}(1)$ Higgs quartic couplings
and ${\cal O}(1)$ new Yukawa coupling $\rho_{tt}$ (and $\rho_{tc}$).
The ${\cal O}(1)$ new Yukawa couplings are understood as
modulo flavor organization as in SM: 
much smaller Yukawa couplings when lower generation is involved.
Much new FPCP --- Flavor Physics and CPV --- phenomenology is implied,
but Nature skillfully hides them, \emph{by alignment}.

We find that alignment naturally emerges with 
${\cal O}(1)$ Higgs quartics in the model.
This is in strong contrast with 2HDM II, 
which is hampered by the $\tan\beta$ parameter
that is unphysical without the $Z_2$ symmetry.
We have illustrated the vast possibilities in parameter space:
extreme alignment is possible if $\eta_6$, which controls $h^0$--$H^0$ mixing,
is of order $m_h^2/v^2 \simeq 0.26$;
but mild alignment, e.g. $\cos\gamma \sim 0.2$ can be
easily entertained with $\eta_1 < \eta_6$ but both being ${\cal O}(1)$,
and $m_h^2$ is pushed down from $\eta_1 v^2$ by level repulsion. 
The ${\cal O}(1)$ nature of Higgs quartics, including $\mu_{22}^2/v^2$, 
suggests that the extra Higgs bosons are sub-TeV in mass
--- a boon to the LHC, 
and very different from the usual projection from SUSY-type 2HDM.
Extra Higgs search therefore needs to be reconsidered.

While alignment by decoupling (in SUSY-type 2HDM) whispers
the thought that exotic Higgs bosons are beyond reach at the LHC
--- we actually might discover them at the LHC.
What would that imply?

\hskip 0.3cm -- It would Not be SUSY!

\hskip 0.3cm -- We could touch EWBG! \quad [hence need CPV probe~\cite{HK}]

\hskip 0.3cm -- Another layer of scale is guaranteed: by the Landau pole(s).
100 TeV collider?

\bigskip
\noindent{\bf Acknowledgement.} 
We thank Kaori Fuyuto, Mariko Kikuchi and Eibun Senaha
 for pleasant and fruitful collaborations,
 and Masaya Kohda for discussions.

\end{document}